\documentclass[10pt]{article}

\usepackage{graphics}
\usepackage{graphicx}

\usepackage[a4paper, left=35mm,right=35mm,top=34mm,bottom=34mm]{geometry}
\usepackage[utf8]{inputenc}
\usepackage[T1]{fontenc}
\usepackage[english]{babel}

\usepackage{enumerate}
\usepackage{graphicx}
\usepackage{hyperref}
\hypersetup{
    colorlinks=true,
    linkcolor=blue,
    filecolor=magenta,      
    urlcolor=cyan,
}
\usepackage{listings}
\usepackage{color}

\definecolor{dkgreen}{rgb}{0,0.6,0}
\definecolor{gray}{rgb}{0.5,0.5,0.5}
\definecolor{mauve}{rgb}{0.58,0,0.82}
\lstdefinelanguage{MRGC++}{%
  language=C++,
  morekeywords={T, U, MPI_Irecv, MPI_Isend, MPI_Allreduce, MPI_Waitall, Compute, Map, abs, max, Swap, MPI_Recv_init, MPI_Send_init, MPI_Startall, Copy, Init, InitRecv, InitSend, InitAllReduce, Send, Recv, AllReduce, Finalize, InitSnapshot, Snapshot, SwitchAsync, SnapReduce, MPI_Test, MPI_Start}
}
\usepackage{mathtools,amsthm,amssymb,amsfonts}
\usepackage{algorithm}
\usepackage{algorithmic}
\makeatother
\theoremstyle{plain}

\theoremstyle{definition}

\theoremstyle{remark}

\usepackage{caption} 
\usepackage{fancyhdr}

\author{
  {\normalsize Guillaume Gbikpi-Benissan}\thanks{Ecole Centrale Paris, France
    (correspondence, frederic.magoules@hotmail.com).}
  \and
  {\normalsize Fr\'ed\'eric Magoul\`es}\footnotemark[1]
}
\title{Beam-tracing domain decomposition method for urban acoustic pollution}
\date{}

\begin{document}
\maketitle
\thispagestyle{fancy}

\begin{abstract}
\noindent This paper covers the fast solution of large acoustic problems on low-resources parallel platforms. A domain decomposition method is coupled with a dynamic load balancing scheme to efficiently accelerate a geometrical acoustic method. The geometrical method studied implements a beam-tracing method where intersections are handled as in a ray-tracing method. Beyond the distribution of the global processing upon multiple sub-domains, a second parallelization level is operated by means of multi-threading and shared memory mechanisms.

Numerical experiments show that this method allows to handle large scale open domains for parallel computing purposes on few machines. Urban acoustic pollution arrising from car traffic was simulated on a large model of the Shinjuku district of Tokyo, Japan. The good speed-up results illustrate the performance of this new domain decomposition method.
\end{abstract}


\section{Introduction}

Important material resources is often a requirement, either for the processing of large volumes of data, or for the fast resolution of large problems. Especially in parallel computing, the effectiveness of the designed algorithms is related to their scaling behavior on massively parallel platforms. However, for a great part of the scientific community, there remains a need for big simulations with very low resources. Then, it could be of major interest to design algorithms with good acceleration properties on very small clusters of machines.

In this study, we tackle the simulation of car traffic noise level at a whole district scale. In a lot of countries, environmental noise has become a major source of stress and of various diseases like hypertension or heart ischaemia. Fast noise simulation could be an effective solution to assess the effect of architectural configurations, in order to reduce the acoustic pollution or to maintain a maximum noise level. Given that sound propagates far away in open domains, a large scale model is required, and well designed methods are a key point to obtain fast simulations.

The more physically correct algorithms are based on numerical methods, such as the Boundary Element Method~\cite{ZZZ2005}, the Infinite Element Method~\cite{magoules:journal-auth:26,magoules:journal-auth:19}, the Finite Element Method~\cite{Ihl1998,Tho2006}, the Stabilized Finite Element~\cite{magoules:journal-auth:7} and the coupling methods~\cite{magoules:journal-auth:15}. Even if these methods accurately approximate the mathematical equations of the acoustic problems, they unfortunately need a lot of computational power and memory. Hence, they are not good candidates for our context of low resources.
Furthermore, relatively to the wavelength, these methods have some difficulties to handle extremely large domains and open domains, such as ones involved in exterior acoustic problems.

Contrariwise, geometrical methods, such as the image-source method~\cite{AB1979,PLS2002}, the ray-tracing method~\cite{Elo2005} and the beam-tracing method~\cite{FTC2004,LSLS2009}, are well suited for large domains and open domains. But only simulations of high frequencies, relatively to the size of the problem, provide accurate results, due to the fact that, in such methods, the sound propagates in straight lines. Fortunately, this is not a limitation for our car traffic noise simulation.

In this study, we consider geometrical methods for the fast simulation of urban acoustic pollution within a large open area. After a brief overview of geometrical methods, we describe a hybrid method coupling beam-tracing and ray-tracing methods, with some similarities to frustum-tracing~\cite{CLTRM2008}. Then we present the parallelization of the algorithm by means of techniques from domain decomposition methods~\cite{MR2006}, widely used for numerical methods in acoustics~\cite{MMC2000}. Finally we illustrate the efficiency of our parallelization technique, using a few machines for simulations run on a model of the shinjuku district of Tokyo, Japan.

\section{Hybrid geometrical acoustic method}

Commonly, geometrical methods compute multiple paths followed by the sound emitted from a source, and measure the acoustic pressure at some points called microphones. The image-source method~\cite{AB1979,PLS2002} creates virtual sources each time a sound ray reflects on a surface inside the 3D model. Sound pressure level is evaluated at a microphone by adding the contribution of the sources and virtual sources for which there is no obstacle on the direct path toward that microphone.

The ray-tracing method~\cite{Elo2005} divides the energy of each source between a huge number of elementary particles (similar to their physical counterpart, the photons) which propagate as sound rays. The energy of a particle gradually decline due to air damping, and it also looses energy when it reflects on a surface. Once this energy reaches a threshold low value, the particle is deleted.

The beam-tracing method is based on a similar principle, except that sound rays are replaced by sound beams, then the source energy is split between beams, according to its power in each direction. This method is harder to implement, mainly because intersections between beams and surfaces are more complex to detect and to handle. For example, each time a beam partially hits one single surface, it needs to be split into sub-beams, what can dramatically increase the complexity of the computation, due to the recursion of this principle. However, for the same simulation quality, there is much less beams to launch, relatively to ray-tracing.

Ray-tracing methods have been extensively studied in the context of computer generated imagery. Efficient generation of hierarchical partitioning of good quality~\cite{SF1992,Wal2004} for the handling of intersection detection, the use of vectorial instructions~\cite{WS2001,Wal2004}, and graphics processing unit (GPU)~\cite{GPSS2007,HSHH2007} allowed to accelerate the ray-tracing. Large scale models got some attention too~\cite{WS2001,CZL2005,LYTM2008}. However, these optimized methods are very efficient when the rays follow similar trajectories, which is rarely the case in acoustic analysis. On the other hand, even if beam-tracing generates great quality soft shadows and anti-aliasing in computer graphics~\cite{ORM2007}, the additional complexity and slowdown do not worth it.

The hybrid approach considered here applies beam-tracing principles, except that the intersections between beams and objects surface are handled as in ray-tracing method. In other words, instead of testing the whole volume of the beam against the model, only the guiding ray of the beam is tested. Hence, in this method, the beams are never split on reflection. This trade-off between precision and speed allows to use very efficient ray shooting methods. In practice, the drawback due to the approximation error can sufficiently be mitigated by processing very small beams. Moreover, a beam becomes larger far from the energy source, where its energy is significantly depleted, what would reduce the impact of an approximation error.

\section{Parallel computing}

\paragraph{Advantage of domain decomposition}

In a standard implementation, the input and the output of parallel computing are sequential. According to the Amdahl's law, the global speed-up of a parallel program containing a percentage $f$ of sequential part is bounded by $1/(1-f)$. One of the main goals of the Domain Decomposition Methods (DDM) is to allow the whole program to be parallel. The idea is to split the model into sub-domains, such that the loading, output and result gathering could be done in parallel for each sub-domain. The splitting of the model can be done only once for multiple simulations.

Another important advantage of the DDM is to allow to consider very large models and a great number of microphones. This is particularly important when generating volumetric noise map, since the number of microphones quickly becomes very large. Each process only needs to allocate memory for the sub-domains it is working on instead of allocating memory for the whole domain.

\paragraph{Microphones partitioning}

In this kind of decomposition, only the set of microphones is split, while the complete geometry is replicated. Each sub-domain is totally independent so the program can be run in parallel without synchronization. Even if a whole part of the work is replicated, especially the intersections detection and the loading/preprocessing of the model, such a decomposition can be efficient for a simple geometry model populated with a huge set of microphones. The absence of communication between the processes is a major advantage in the context of a largely distributed or loosely connected platform. However, there is no easy way to mitigate load balancing issues when the processing time of sub-domains varies a lot.

\paragraph{On demand geometry and microphones loading}

With prior computation of some hierarchical acceleration structure, sub-domains can be loaded only when needed, but the actual parallelization is done on the set of beams. At the beginning, first level of the hierarchy is loaded, and throughout the computation, this structure is accessed several times. If during the traversal of the structure, a node not yet loaded is reached, the corresponding data are loaded. If it is an interior node, these data are the child nodes information, if it is a leaf, the data are the corresponding mesh and microphone data. Beside the specific precomputed acceleration structure, this model of parallelization is quite hard to implement. The latency for loading a node of the hierarchical structure should be hidden, for instance by changing the current beam to one whose data are ready. The gathering of the final results is also complicated since the data are spread over all processes, and potential copies need to be merged. The main drawback is the risk for each process to load the complete model when beams largely spread inside the area. That is particularly the case in exterior acoustics problems where beams often go far and are widely dispersed.

\paragraph{Geometry and microphones partitioning}

The original method used in this work matches more closely a domain decomposition approach~\cite{MR2006}. Here, both the geometry and the microphones are split into multiple sub-domains. When a beam goes out of a sub-domain, it is sent to the process working on the sub-domain intersected. Interface conditions~\cite{magoules:journal-auth:16} are used to assure the continuity of the beam properties from one sub-domain to another one. Efficient interface conditions can be designed by a continuous approach~\cite{Des1993,Gha1997,CN1998}.
Continuous optimized approaches improve the convergence of the algorithm~\cite{magoules:journal-auth:6,magoules:journal-auth:28,magoules:journal-auth:23,magoules:journal-auth:18,magoules:journal-auth:14,magoules:journal-auth:28}.
Similarly, the performance of the algorithm can be increased by using a discrete optimized approach~\cite{magoules:journal-auth:8,magoules:journal-auth:29,magoules:journal-auth:20,magoules:journal-auth:17,magoules:journal-auth:12,magoules:proceedings-auth:6}.
In~\cite{magoules:journal-auth:30}, a link is established between these two approaches.
In this paper we consider a domain decomposition approach as described in~\cite{magoules:journal-auth:10,magoules:journal-auth:9,magoules:journal-auth:13} but extended to beam-tracing~\cite{magoules:patent:2011}.
In contrast, unlike classical domain decomposition methods, load-balancing issues can not be fixed in a static way, where a one-to-one correspondence is done between the processes and the sub-domains. Indeed, considering for instance the case where there would be only one sound source, the sub-domain containing the source would have the most computational load. This is why a more complex load balancing scheme has to be used. The idea is that each process starts with one sub-domain, but when it has few remaining beams to handle, it starts loading one or more sub-domains still containing a lot of unhandled beams. The number of sub-domains which can be loaded simultaneously is limited by the memory size. In order to overlap the results gathering time, sub-domains can also be unloaded when few beams remain to be processed. Since the gathering operation mainly uses the communication system, this overlapping does not slow down the computation. Nonetheless, the loading and unloading of sub-domains take a long time, hence it would be preferable to avoid such operations as much as possible. For that, it is still relevant to approximate at best an average good load balancing. The solution used is to process first the exchanged beams and then to equalize the number of remaining beams in each sub-domain comparatively to the processing power affected to this sub-domain by loading or unloading sub-domains only when this ratio deviates too much from the average.

At last, the processing of a sub-domain itself is parallelized by coupling a shared memory multi-threading and a work-sharing load balancing. This allows an additional acceleration of the computation without any data replication. There is no need of gathering local outputs at the end since the output data are shared. However, it required to implement a thread-safe access mechanism. A simple fine-grained locking were implemented by setting a spin-lock mutex on each cell of the output array. Given that the multi-threaded beam-tracing generates few contention, the busy-waiting time is insignificant, hence the spin-lock is an acceptable trade-off between simplicity and efficiency for our acoustic problem.

\section{Results and discussions}

\begin{figure}[h!]
  \begin{center}
    \includegraphics[width=0.75\textwidth]{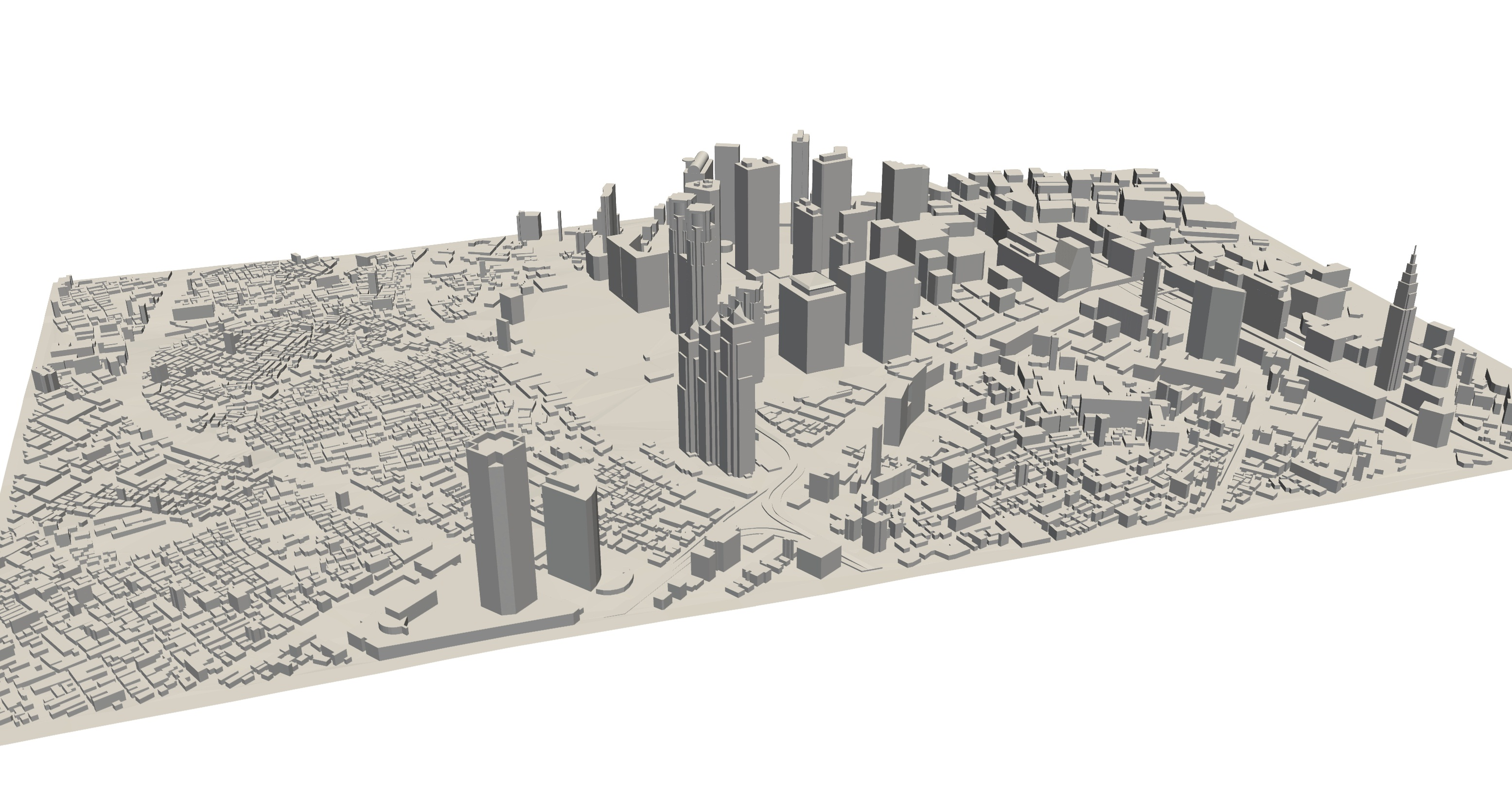}
  \end{center}
  \caption{Global view of the virtual model of Shinjuku, district of Tokyo, Japan.} 
  \label{fig:cad1}
\end{figure}

\begin{figure}[h!]
  \begin{center}
    \includegraphics[width=0.645\textwidth]{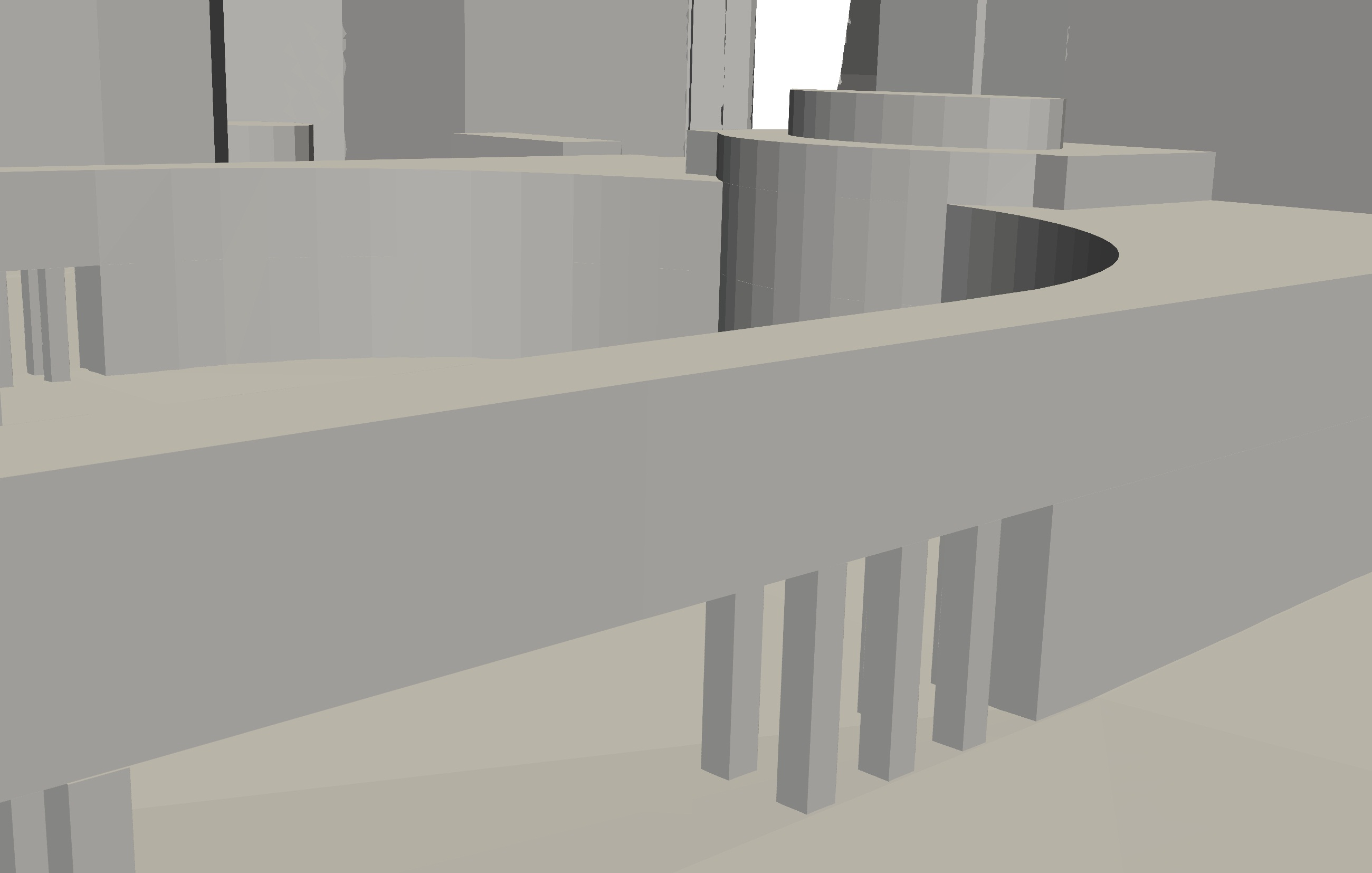}
  \end{center}
  \caption{Close view of the virtual model of Shinjuku, just next to the Tokyo Metropolitan Government Building.} 
  \label{fig:cad2}
\end{figure}

\begin{figure}[h!]
  \begin{center}
    \includegraphics[width=0.645\textwidth]{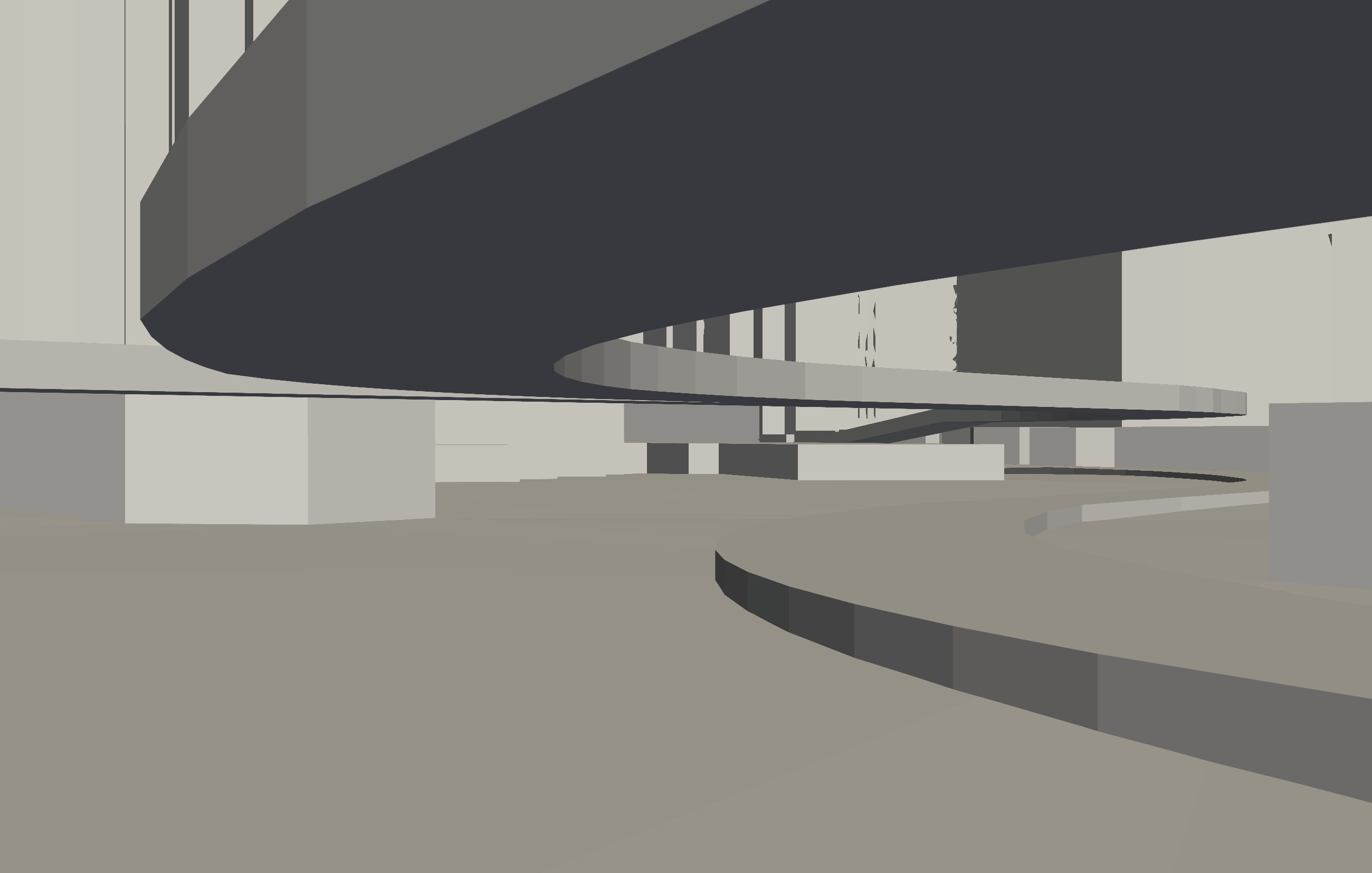}
  \end{center}
  \caption{Close view of the virtual model of Shinjuku, under the Metropolitan Expressway No. 4.} 
  \label{fig:cad3}
\end{figure}

In this study, we tackled the simulation of urban acoustic pollution related to car traffic. Experiments were run on a model of the Shinjuku district of Tokyo, Japan. It is an area of 2.5 km by 1.5 km around the point of coordinates $35^{o}41^{'}23^{''}N$, $139^{o}1^{'}25^{''}E$. By hosting the busiest train station of the world, multiple commercial and administrative buildings including the administration center of the government of Tokyo, and around ten skyscrapers at least 200 meters high, the Shinjuku district faces a large number of activities generating a very important car traffic. Figure~\ref{fig:cad1} shows a global view of the simulation model. The whole district buildings has been represented, and as we can see on figures~\ref{fig:cad2} and~\ref{fig:cad3}, it is a detailed designing leading to a quite large model.

\begin{figure}[h!]
  \begin{center}
    \includegraphics[width=0.75\textwidth]{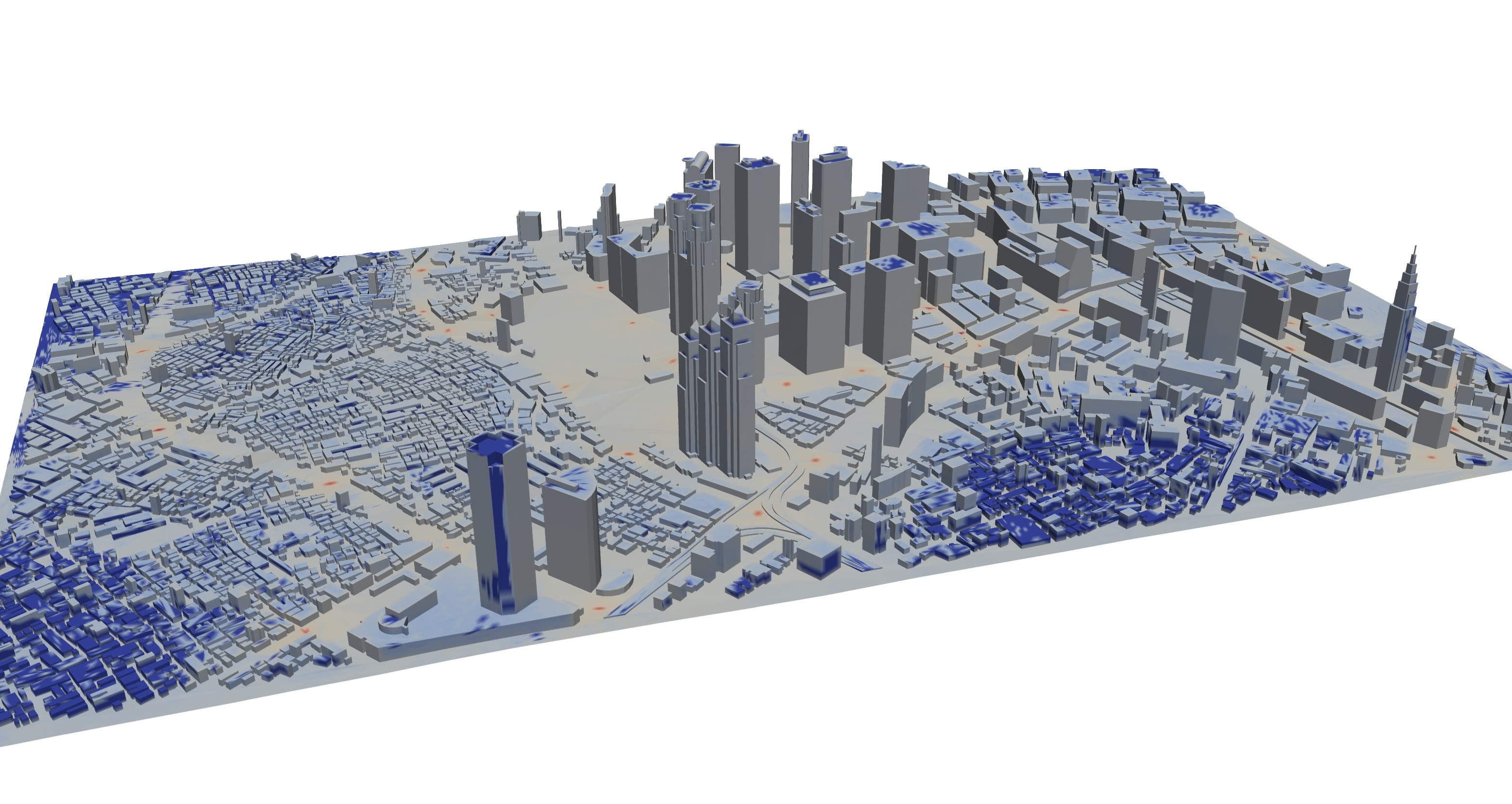}
  \end{center}
  \caption{Simulation of a sound pressure level distribution in Shinjuku, district of Tokyo, Japan.} 
  \label{fig:spl}
\end{figure}

As we were saying at the beginning, our goal is to study the efficiency of the proposed DDM-based hybrid geometrical method on low-computer resources architectures. The numerical experiments were run on a hybrid, both distributed and shared memory, computational platform consisting of 8 workstations, each one containing two quad core processors (a total of 64 cores). Each machine was provided with 8 Gigabytes RAM (Random Access Memory) and a Windows 7 operating system. The volume of the hexahedral bounding box of the model was populated with a total number of $13.7$ millions of microphones, regularly distributed on a 3D grid. According to the dimensions of the district, that corresponds to a microphone every 4 meters in each of the three spatial directions. We simulated 49 sound sources, most of them placed at major crossroads, and each one launching 3 millions of beams. Figure~\ref{fig:spl} illustrates the noise level distribution obtained from such a simulation.

Speed-ups are computed using the total execution time, including the loading of the model, the acceleration structure generation and the saving of the results. A speed-up $S$ is determined by $S = {t_{ref}}/{t}$, where $t_{ref}$ is the execution time of the sequential version, and $t$, the time consumed by the parallel version. Table~\ref{tab:ddm_acoustic} shows different speed-ups relative to various DDM/multi-threading schemes.
\begin{table}[h!]
\caption{Speed-up of the acoustic DDM program (Ethernet) with respect to the number of threads and sub-domains.}
\label{tab:ddm_acoustic}
\centering
{\small
\begin{tabular}{|l|c|c|c|c|}
\hline 
 & 16 & 32 & 64 & 128 \\
 & threads & threads & threads & threads \\
\hline %
{1 sub-domain}  & 16.03 & 25.20 & 30.26 & 23.19 \\
{2 sub-domains} & 16.28 & 30.64 & 49.25 & 56.64 \\
{4 sub-domains} & 16.15 & 29.36 & 50.82 & 54.26 \\
{8 sub-domains} & 16.24 & 29.16 & 52.79 & 83.72 \\
\hline 
\end{tabular}
}
\end{table}
The computation was distributed such that each core runs exactly 2 threads, which corresponds to 16 threads per workstation. Then, for the 16, 32, 64 and 128 threads, we used respectively 1, 2, 4 and 8 workstations. Even if we are not in a scaling study, one could remark that the classical parallelization without DDM underperforms from 64 threads (4 nodes) to 128 threads (8 nodes). The use of multiple sub-domains allows a bigger part of the application to be parallelized, and gives better results. Let us consider a sequential version consuming about 4 hours to complete the whole simulation. With only 4 workstations, the simple multi-threaded simulation would last about 8 minutes (speed-up for 1 sub-domain and 64 threads), while the version with 8 sub-domains would not exceed 5 minutes. However, that is the limit for the classical hybrid distributed/shared-memory parallelization. According to the last result in the table, the DDM approach would allow to reduce the maximum execution time from 4 hours to 3 minutes, using 8 workstations composed of 2 quad-core processors with 8 GB RAM. Table~\ref{tab:ddm_acoustic_time} shows parallel execution times (in seconds) for our simulation where the sequential processing lasted 5392 seconds (1 hour and 30 minutes).
\begin{table}[h!]
\caption{Execution time (in seconds) of the acoustic DDM program (Ethernet) with respect to the number of threads and sub-domains.}
\label{tab:ddm_acoustic_time}
\centering
{\small
\begin{tabular}{|l|c|c|c|c|}
\hline 
 & 16 & 32 & 64 & 128 \\
 & threads & threads & threads & threads \\
\hline %
{1 sub-domain}  & 336,31 & 213,96 & 178,18 & 232,53 \\
{2 sub-domains} & 331,24 & 175,98 & 109,47 & 95,20 \\
{4 sub-domains} & 333,97 & 183,68 & 106,09 & 99,37 \\
{8 sub-domains} & 332,11 & 184,93 & 102,14 & 64,40 \\
\hline 
\end{tabular}
}
\end{table}

\section{Conclusion}

Acoustic simulation software is an invaluable tool to handle noise problems, both inside closed an open areas. In this paper, we reminded some principal geometrical methods which basically consider the sound propagation as straight line paths shot from a sound source. An hybrid beam/ray-tracing algorithm was parallelized by means of domain decomposition methods coupled with a dynamic load balancing. An additional acceleration was gained by a second level parallelization based on multi-threading in shared memory environment.

Numerical simulations have been carried out to evaluate the performance and efficiency of this new method on low-resources platforms, for analyzing large scale urban acoustic pollution. This new method allows to obtain a significant acceleration on few processors by parallelizing the whole algorithm, including the gathering process. As our experiments show, it can be used on inexpensive systems, such as a cluster of workstations provided with an Ethernet interconnection, since it reduces the memory usage and the communication bandwidth.

\bibliography{ref}
\bibliographystyle{abbrv}

\end{document}